\documentclass[12pt,article]{revtex4}
\usepackage{color}
\usepackage{graphicx}
\usepackage{amsmath}
\usepackage{amssymb}
\usepackage{epsfig}
\usepackage{wasysym}
\usepackage{bbm}
\usepackage{multirow}
\renewcommand{\narrowtext}{\begin{multicols}{2} \global\columnwidth20.5pc}
\renewcommand{\v}[1]{{\bf #1}}

\newcommand{\s}{{\sigma}}

\def\be{\begin{eqnarray}}
\def\ee{\end{eqnarray}}

\newcommand{\nn}{\nonumber\\}

\newcommand{\Eq}[1]{Eq.~(\ref{#1})}

\newcommand{\ra}{\rightarrow}

\newcommand{\e}{\epsilon}

\newcommand \ti[1]{}

\begin{document}

\title{Concepts relating magnetic interactions, intertwined electronic orders
and strongly correlated superconductivity}
\author{J.C. Davis} \affiliation{Department of Physics, Cornell University, Ithaca, NY 14853, USA.}
 \affiliation{CMPMS Department, Brookhaven National Laboratory, Upton, NY 11973, USA.}
 \affiliation{School of Physics and Astronomy, University of St. Andrews, Fife KY16 9SS, Scotland.}
 \affiliation{Kavli Institute at Cornell for Nanoscale Science, Cornell University, Ithaca, NY 14853, USA.} \author{D.-H. Lee} \affiliation{Department of Physics, University of California, Berkeley, CA, 94720, USA.}  \affiliation{Materials Science Division, Lawrence Berkeley National Lab, Berkeley, CA, 94720, USA.}



\begin{abstract}
Unconventional superconductivity (SC) is said to occur when Cooper pair formation is dominated by repulsive electron-electron interactions, so that the symmetry of the pair wavefunction is other than isotropic s-wave.  The strong, on-site, repulsive electron-electron interactions that are the proximate cause of such superconductivity are more typically drivers of commensurate magnetism.  Indeed, it is the suppression of commensurate antiferromagnetism (AF) that usually allows this type of unconventional superconductivity to emerge. Importantly, however, intervening between these AF and SC phases, ``intertwined'' electronic ordered phases of an unexpected nature are frequently discovered. For this reason, it has been extremely difficult to distinguish the microscopic essence of the correlated superconductivity from the often spectacular phenomenology of the intertwined phases. Here we introduce a model conceptual framework within which to understand the relationship between antiferromagnetic electron-electron interactions, intertwined ordered phases and correlated superconductivity. We demonstrate its effectiveness in simultaneously explaining the consequences of antiferromagnetic interactions for the copper-based, iron-based and heavy-fermion superconductors, as well as for their quite distinct intertwined phases.
\end{abstract}
\maketitle
\section{INTRODUCTION}
 Emergence, the coming into being through evolution, is an important concept in modern condensed matter physics\cite{Anderson}. Superconductivity is a classic example of emergence in the realm of quantum matter: as the energy-scale decreases, the effective electron-electron interactions responsible for Cooper pairing and thus superconductivity evolves from the elementary microscopic Hamiltonian through unanticipated modifications\cite{Schrieffer}. This is why it is so difficult to derive superconductivity from first principles. Finding the microscopic mechanism of Cooper pairing means discovering the nature of the ultimate effective electron-electron interaction at the lowest energy scales.

In the last three decades, unconventional\cite{Monthoux, Norman, Scalapino} forms of superconductivity have been discovered in many strongly correlated (repulsive electron-electron interaction) systems. These materials fascinate a lay person for their high superconducting transition temperatures, and therefore the potential for revolutionary applications in power generation/transmission, transport, information technology, science, and medicine.  They intrigue (and challenge) physicists to identify the mechanism of their high pairing-energy scale and because of the many ``intertwined''\cite{Berg,Fk} electronic phases that have been discovered in juxtaposition with the unconventional superconductivity. These have been hypothesized to ``arise together from one parent state'' such that ``the various order parameters are intertwined rather than simply competing with each other''\cite{Fk}. The best known and most widely studied examples include the copper-based\cite{Orenstein, Shen, Pepin, PALee}  and iron-based\cite{Paglione,Wang,Hirschfeld} high temperature superconductors,  the heavy-fermion superconductors\cite{Heffner,Pfleiderer,Coleman}  and the organic superconductors\cite{Saito} . One thing commonly noted in these systems is that superconductivity normally borders antiferromagnetism: in the phase diagram spanned by temperature and a certain control parameter (chemical-doping, pressure...etc.), a superconducting (SC) dome stands adjacent to the antiferromagnetic (AF) phase (Fig.1). However, the precise way the two phases are connected varies greatly from system to system.

Another very common observation is the appearance of other ordered phases of electronic matter that ``intertwine'' with the superconductivity. These exotic intertwined phases (IP) occur in the terra incognita between the superconductivity and the antiferromagnetism (gray Fig.1). Examples include the charge/spin density wave\cite{Kivelson,MFujita,KFujita} and intra-unit-cell symmetry breaking\cite{KFujita,Bourges,Lawler}  orders in the copper-based superconductors, and the nematic order\cite{Chuang,Chu} in the iron-based superconductors. 
A key  long-term objective for this field has therefore been to identify a simple framework within which to consider the relationship between the antiferromagnetic interactions, the intertwined electronic orders that appear at its suppression, and the correlated superconductivity.

Because in all the systems considered here superconductivity emerges from the extinction of antiferromagnetism, it is widely believed that the effective electron-electron interaction triggering the Cooper pairing could be antiferromagnetic in form. In that case, of course, the same argument could apply to the other intertwined electronic phases. These ideas motivate the assertion that  antiferromagnetic effective electron-electron interactions may drive both the correlated superconductivity and the other intertwined phases. Until recently, however, there has been little  consensus on this issue. One reason is that the experimental evidence for many such intertwined states has only been firmly established in recent years. Another reason is that while magnetism in proximity to unconventional superconductivity appears universal, the nature of the intertwined phases changes from system to system  for reasons that appear mysterious. 


In this paper, we therefore explore the plausibility that an antiferromagnetic effective interaction could be  the driving force for {\it both} the unconventional superconductivity and the intertwined orders in the copper-based, iron-based and heavy-fermion, superconductors. (We omit discussion of organic superconductors, see, e.g., Ref.\cite{Taillefer} and Ref.\cite{Kanoda}, for the sake of brevity.) Here we will not try to rigorously solve for the ground state under different conditions. Our goal is to ask whether the known intertwined states are the locally stable mean-field phases when the sole effective electron-electron interaction is antiferromagnetic. We understand that the actual effective interactions may be more complex than this simplest antiferromagnetic form; we deliberately omit these details with the goal of identifying a simple framework within which all the relevant phenomena can be considered. 
Two very recent  papers based on a related approach, but focusing only on the copper oxide superconductors, have appeared\cite{Sachdev2,Laughlin}.

\section{The effective Hamiltonian}

Thus we start by studying the assertion that fermiology (the Fermi surface topology) + antiferromagnetic \underline{effective} electron-electron interaction can generate the known intertwined phases in different types of correlated superconducting materials. Our effective Hamiltonian is viewed as evolved from the bare Hamiltonian for strong Coulomb interactions, and our strategy is to explore, in different ordering channels, which order dominates as the exchange constants of the effective interactions increase from zero. Under these circumstances, it is the antiferromagnetic interaction that is universal while it is the fermiology that is not. 
The effective Hamiltonian corresponding to our assertion is
\be
H_{eff}={\sum_{\v k}}'\sum_{s}\e(\v k) n_{s}(\v k)+\sum_{i,j}~J_{ij}~\v S_i\cdot\v S_j.\label{eff}\ee
Here $\v k, s$ are the momentum and the spin labels, respectively. $\v S_i$ represents the total spin operator in the i$^{\rm th}$ unit cell, it is given by ${1\over 2}\sum_{\tau,s} c^\dagger_{i\tau s}\vec{\s}_{ss'} c_{i\tau s}$, where $\tau$  labels the degrees of freedom in each unit cell (e.g., orbital, different sites ...etc). In addition, \Eq{eff}  ${\sum}'_{\v k}$  is a sum restricted to the neighborhood of the Fermi surface  (which can have several disconnected pieces), $\e(\v k)$ is the dispersion of the relevant band in the vicinity of the Fermi surface, and  $J_{ij}~\v S_i\cdot\v S_j$ should be understood as a electron-electron scattering term. Although we write it in real space, it should be converted to the band eigen-basis and projected to the neighborhood of the Fermi surface, for each different type of system.

Obviously, many simplifying assumptions have already been made here. Note that, aside from the fact that ${\sum}'_{\v k}$ restricts states to low single-particle excitation energies, there is no further constraint on the Hilbert space, there is no gauge field, and the particle statistics are the usual Fermi statistics. The only effect of interactions is captured by the $J_{ij}$ term. This asserts that the low energy physics, even non-fermi liquid behavior, can be the result of the 
antiferromagnetic effective interaction. Therefore although in the absence of $J_{ij}$ \Eq{eff} describes a Fermi liquid, in its presence the system may behave otherwise 
precisely because $J_{ij}$ can drive many intertwined instabilities, and strong (critical) fluctuations between these instabilities can then drive non-fermi liquid behavior.  Thus writing down \Eq{eff} is not equivalent to assuming an ``nearly antiferromagnetic Fermi liquid''\cite{Pines}. This is particularly so near, e.g., the antiferromagnetic quantum critical point where $J_{ij}$ can exhibit a strong dependence on the energy cutoff down to the lowest energy.

 Of course, we do understand that many learned readers may question our starting point of \Eq{eff}. However, in the search for a simple conceptual framework within which to understand quite different correlated superconductors along with their distinct and complex intertwined phases in multiple material systems\cite{Berg,Orenstein,Shen, Pepin, PALee,Paglione,Wang,Hirschfeld}, such a simple starting point  can have many advantages.

\section{The Copper based superconductors}


For the case of the copper-based superconductors\cite{Berg,Orenstein,Shen, Pepin, PALee}, we use a simple one-band model to describe the first term of \Eq{eff}; the relevant Fermi surface is shown in Fig. 2(a). 
And for $J_{ij}$ we use the simplest nearest neighbor interaction to emulate the antiferromagnetic correlations.
The utility of \Eq{eff} is validated in part by the Fermi liquid quasiparticle Landau quantization observed by high field quantum oscillation experiments\cite{Taillefer2,Sabastian}; it is theoretically plausible\cite{Senthil} that such Fermi liquid behavior can be  regained when the strong magnetic field quenches the relevant fluctuations.

It has been known since the early days of cuprate superconductivity that antiferromagnetic fluctuations can induce d-wave Cooper pairing\cite{Loh,Rice,Kotliar}. We begin by reproducing what is known. Using the effective Hamiltonian specified above we obtain (Methods Section I) the leading and sub-leading superconducting gap functions shown in Fig. 2(c) and Fig. 2(d). (The idea to mean-field decouple the magnetic interaction to obtain Cooper pairing originates from Ref.\cite{rvb,BZA}.)   These two gap functions are approximately described by $\cos k_x -\cos k_y$ and $\cos k_x +\cos k_y$ respectively. This indicates that the cuprates can have extended s-wave pairing tendency\cite{BZA} after all. Furthermore,  Fig.2(b) shows how this Fermi surface exhibits eight special ``hot spots'' (the red dots) where the AF Brillouin zone (dashed lines) crosses it. These are hypothesized to play a leading role in the interplay of intertwined phases and superconductivity in cuprates; see e.g. Ref.\cite{Sachdev2}.

In the particle-hole interaction channel, \Eq{eff} predicts (Methods Section II) two types of instabilities:
one preserves translation invariance (a $\v Q = 0$ instability) and the other (a finite $\v Q$ instability)
does not. Within our approach, the leading $\v Q = 0$
instability is to a nematic state. The order parameter and the associated Fermi surface distortion are shown in Fig. 3(a).
This instability leads to the breaking of the crystal $90^\circ$ rotation symmetry which has been reported within the CuO$_2$
unit cell\cite{KFujita}. The fact that the cuprate Fermi
surface has such a tendency to Pomeranchuk distort has been widely discussed (see, e.g., Ref.\cite{Fradkin}).

 The subleading Q = 0 instability is two fold
degenerate. The order parameters and the associated Fermi surface distortions are shown in Fig.3(b) and 3(c). (Similar instabilities in hexagonal systems were discussed recently in Ref.\cite{Maharaj}.) Because the distorted Fermi surfaces are not $\v k\leftrightarrow -\v k$ symmetric, these instabilities lead to time reversal symmetry breaking. In Fig.4(a) we show the ground state current distribution produced by these order parameters in Fig.3(b,c). Depending on the quartic term in the Landau free energy expansion the order parameters of panel Fig.3(b,c) can coexist. In Fig.4(b) we show the  ground state intr-unit-cell currents associated with the symmetric and antisymmetric combination of the order parameters in Fig.3(b) and Fig.3(c) respectively. Clearly this subleading time reversal breaking Pomeranchuk instability leads to states with the same broken symmetry as the loop current states proposed in Ref.\cite{varma}, and not inconsistent with reported time reversal symmetry breaking in cuprates\cite{Kaminski,Fauque,Xia}. However, it is important to stress that our $\v Q=0$ instability does not lead to a pseudogap. Moreover, although this instability is subleading here, it is possible that material dependent details omitted in our simple effective action can change that.


The leading $\v Q\ne 0$ instability in the particle-hole channel is a charge density wave (CDW) instability (in a recent preprint\cite{Sachdev2} a related idea was discussed). The subject of CDW order in cuprate superconductors has a long history. An apparently bi-directional modulated CDW with only short range order is widely observed using spectroscopic imaging scanning tunneling microscopy\cite{KFujita} but it was difficult to be certain these were true bulk phenomena. Therefore, for a long time the only bulk charge density wave order that was firmly established experimentally was the unidirectional charge density wave (stripes)\cite{Kivelson} in the La$_2$BaCuO$_4$ family of compounds\cite{MFujita}. Recently, however,  signatures of apparent bidirectional CDW order have been observed by X-ray scattering in bulk YBa$_2$Cu$_3$O$_7$-crystals\cite{Hayden,Keimer,Li}.

In Fig.5(a,b) and Fig.5 (c,d) we present the leading charge density wave order parameters
that  are generated by \Eq{eff} (Methods Section II). 
Panels Fig.5(a,b) and Fig.5(c,d) represent the CDW order parameters whose ordering wavevectors are the four horizontal
and vertical black arrows connecting the ``hot spots'' in Fig.2(b).  (The gray arrows are the ordering wavevector of the subleading
charge density wave order that we find (not shown). This is different from the result of Ref.\cite{Sachdev2} where the gray arrows are the leading CDW wavevectors, perhaps due to the difference in the details of effective interaction and bandstructure used in the two approaches.) 
At the quadratic level in a Landau free energy expansion the order
parameters in panel 5(a,b) are degenerate with those in panels 5(c,d). Depending on the coefficients
of the fourth order terms they can be either mutually exclusive (which results in unidirectional
charge density wave) or coexist (which results in bi-directional charge density wave). In Fig.5(e)
we show the energy gap of a bi-directional charge density wave which corresponds to the out-of-phase
coexistence of the order parameters in Fig.5(a-d)\cite{Li}. 

Finally, the fact that there are both strong nematic (Fig. 3(a)) and charge density wave (Fig. 5(a-d)) susceptibilities
implies that, in the presence of disorder, which can serve as localized external ordering fields,
locally nematic and charge density wave ordering can be induced to coexist. This is consistent with the STM
experiments\cite{KFujita}. Such short-range disordered induced ordering can exist even when in the clean
limit the system is not yet long-range ordered.

Obviously there is another key issue requiring discussion here -- the pseudogap of the cuprates. This unexplained gap to single-electron excitations is anisotropic in $\v k$-space and appears at $T^*>>T_c$ for underdoped cuprates\cite{Norman,Orenstein,Shen,PALee,Timusk}. We hypothesize that the consequences of an effective Hamiltonian as described in \Eq{eff} could also account for such a pseudogap. The various instabilities (except those at $\v Q=0$) discussed here can all gap out, at least partially, the single-electron excitation spectrum. However due to the intertwining of these instabilities the order parameter may fluctuate from one type to another. This fluctuation would prevent the system from becoming long range ordered without eliminating the actual "pseudogap" for the single-electron excitations.

Thus we consider the order parameters of different intertwined orders to form a multi-component super-vector. (The notion of super-vector has been discussed in Ref.\cite{Zhang}.)  The magnitude of the super-vector is then responsible for the single-particle gap. The direction of the super-vector is the soft degree of freedom, which ultimately determines the long ranged order of the system. 
 However, while this direction fluctuates the single-electron excitation spectrum remains gapped. In our case, when the gap is partial, the low energy excitations include both the directional fluctuations of the super-vector and the remaining gapless single particle excitations. Of course because of the coupling with the collective excitations, these single-particle excitations can have unusual, e.g., non-fermi liquid, properties.

Now it remains to show with a super-vector formed using the antiferromagnetic, superconducting, and charge density wave order parameters, that there is a pseudogap in the single electron excitation spectrum, no matter where the super-vector points. The results are shown in Fig.6 (Methods Section III and IV).
In Fig. 6(a) the super-vector points in the antiferromagnetic direction. This is shown by the red arrow on the order parameter sphere on the right. Such an order has the biggest effect at the ``hot spots'' (Fig.2(b)) where the gap is maximal (see the left panel of Fig.6(a)). Here and in other panels, a vanishing single-particle gap at any point on the fermi surface means that, along the normal direction there remains Fermi crossing, i.e., the Fermi surface has either moved or reconstructed. The super-vector in Fig. 6(b) points in the superconducting direction. The gap spectrum shown on the left is the familiar d-wave gap. In Fig.6(c) the super-vector points in the charge density wave direction. The energy gap spectrum on the left shows a nodal feature. The super-vector in Fig. 6(d) lies in the plane spanned by the superconducting and the charge density wave but directionally between the two. Finally in Fig. 6(e) the super-vector points in a generic direction. Obviously in reality, different components of the
super-vector do not have to have the same norm so that the fluctuations of the super-vector actually occur on a spheroid, hence there is no enlarged symmetry. We hope this figure makes the heuristic case that a pseudogap can also be a consequence of the effective interaction in \Eq{eff} when the effects of fluctuating intertwined order parameters are dominant. Many of the anomalous physical properties in the pseudogap state could then be attributed to a orientational fluctuations of this intertwined supervector.

\section{The Iron-based superconductors}

Next we carry out the equivalent exercise for the iron-based superconductors\cite{Paglione,Wang, Hirschfeld}. The first term of \Eq{eff} is studied here using a five-band tight-binding model with the Fermi surface shown in Fig.7(a).
The blue and red lines mark the hole and electron Fermi surfaces, respectively. To simulate the magnetic correlation in iron-based superconductors we include both first ($J_1$) and 2nd ($J_2$) neighbor interaction in $J_{ij}$ . (A similar Hamiltonian, with a doped Mott insulator basis, was used to analyze Cooper pairing in pnictides in Ref.\cite{Hu}.) This effective interaction has been derived from the functional renormalization group calculation\cite{Zhai}. Phenomenologically there is mounting evidence that the magnetic correlations in the iron-based materials are not due to fermi surface nesting\cite{Dai}. It is then more appropriate to view the second term in \Eq{eff} as being generated by excitations over the entire bandwidth. The essential difference from a Mott insulator here is the absence of a charge gap. Therefore, the generation of the effective magnetic interactions is more gradual.
 Using these inputs for $J_2/J_1\ge 0.7$ we find (Methods Section I) the leading and sub-leading superconducting order parameter shown in Fig. 7(b) and Fig. 7(c) respectively. The leading gap function has the $S_\pm$ symmetry\cite{Mazin} and the sub-leading one has $d_{x^2-y^2}$ symmetry\cite{Wang,Hirschfeld}.

In the particle-hole interaction channel we find that (Methods Section II) the iron-based superconductors also  have strong $\v Q = 0$ instabilities. In Fig. 8(a) and Fig. 8(b) we show the leading and sub-leading Fermi surface distortions that we determine from \Eq{eff} when the distortion amplitude is small. Here  the un-distorted Fermi surface shown using dashed lines.
This result agrees with the functional renormalization group findings\cite{Zhai}. The leading Fermi surface distortion preserves the point group symmetry of the crystal. Note that because both electron and hole pockets shrink, it preserves the total charge density. (We note that a large amplitude distortion of this type can drive the system to undergo a semi-metal to insulator transition.) The subleading $\v Q=0$ instability breaks the 90$^\circ$ rotation symmetry. Although it is sub-leading at the quadratic level of the Landau free energy expansion, it can become leading once the cubic coupling with the (strong) antiferromagnetic fluctuation is taken into account (note that the antiferromagnetic order in the iron-based materials also breaks the 90$^\circ$ rotation symmetry\cite{PCD}). In Fig.8(c) we show the effect of the symmetry breaking distortion we find on the orbital occupation $n_{xz}(\v k) - n_{yz}(\v k)$. The fact that one needs magnetic fluctuations to stabilize the $C_4$-breaking Fermi surface distortion is consistent with the arguments presented in Ref.\cite{Paglione,Steve,Sachdev3}.
 Thus the result in Fig. 8(b,c) can explain the ubiquitous "nematic" ordering found in the iron-based superconductors\cite{Chuang,Chu}. It also accounts for the photoemission observation of the substantial difference in the d$_{xz}$ and d$_{yz}$ orbital occupation in the nematic distorted state\cite{Yi}.

 Within our approach of \Eq{eff}, iron-pnictides show a very weak $\v Q\ne 0$ CDW instability. The ordering wavevector of the leading CDW is approximately $(\pi,\pi)$. However due to the poor overlaps between the Fermi surfaces upon the $(\pi,\pi)$ displacement, and the fact that $(\pi,\pi)$ only approximately connects electron with electron or hole with hole pockets, a weak CDW will not gap out the fermi surfaces. Therefore we will not devote more space to consideration of the CDW instability in pnictides.

\section{The heavy fermion superconductors}

Finally, we use this same conceptual framework to consider the canonical heavy-fermion superconductor CeCoIn$_5$. The band structure we used for this material is the one given in Ref.\cite{Allan}. The tight-binding model consists of two orbitals per unit cell - the Wannier orbitals associated with the light and heavy band respectively. The Fermi surface is shown in Fig. 9(a). The $J_{ij}$ we use to emulate the antiferromagnetic correlation in CeCoIn$_5$ is the simple nearest neighbor interaction. With these inputs we determine from \Eq{eff} (Methods Section I) the leading and sub-leading superconducting order parameter, the results are shown in Fig. 9(b) and Fig. 9(c).
The leading superconducting gap function, with $d_{x^2-y^2}$ symmetry (Methods Section I), is in excellent agreement with that determined by the STM quasiparticle interference spectroscopy recently\cite{Allan}. The reason that the superconducting gap primarily opens on the large Fermi surface centered at $(\pi,\pi)$ is because the "hot spots" associated with the antiferromagnetic scattering all reside on  that Fermi surface. This is shown in Fig.9(d). Like the cuprates, the subleading superconducting gap function has extended S-wave symmetry.

In the particle-hole channel  our general approach in \Eq{eff} also predicts  that CeCoIn$_5$  has $\v Q = 0$ and $\v Q \ne 0$ instabilities (Methods Section II). The order parameter and the Fermi surface distortion associated with the leading $\v Q = 0$ instability is shown in Fig.10(a). This distortion breaks the crystal $90^\circ$ rotation symmetry and leads to nematicity. It is very interesting that, like the cuprates, the subleading $\v Q=0$ instability is also to a degenerate pair of time reversal symmetry breaking states. The order parameter and the distorted Fermi surfaces are shown in Fig.10(b) and 10(c). The obvious similarity between the $\v Q=0$ instabilities in the heavy fermions and the cuprates is quite striking.

The order parameter of the leading (weak) $\v Q \ne 0$ charge density wave instability is shown in Fig. 11(a-d). The energy gap produced by the in-phase coexistence of the order parameters in Fig.11(a,b) with those in Fig.11(c,d) is shown in Fig.11(e). 
Experimental searches of the signatures of these instabilities are under way.

Searching for instabilities intertwined with superconductivity in heavy fermion compounds now seems an important future direction. However, one must bear in mind that the equivalent chemical pressure places CeCoIn$_5$ near the ``optimal pressure'' where the superconducting transition temperature is the highest. The intertwined instabilities tend to occur near the junction between antiferromagnetism and superconductivity. Therefore unless negative pressure can somehow be applied they can remain out of reach for CeCoIn$_5$. A better system for realizing intertwined instabilities is CeRhIn$_5$ which is antiferromagnetic at ambient pressure. By carefully studying pressure-temperature phase diagram one might be able to find similar phenomena as in underdoped cuprates. If so, this will give additional support for applicability of the simple theory envisioned in \Eq{eff}.

\section{Conclusion}

From the above studies, using the simple concept of the controlling influence of antiferromagnetic electron-electron interactions, it seems fair to say that the low-energy effective Hamiltonian given by \Eq{eff} can be very useful in achieving an elementary understanding of the superconductivity and the intertwined instabilities in several canonical classes of unconventional superconductors. Specifically we note that these studies demonstrate why, while superconductivity is universal, the nature of the Fermi surface distortion and/or the density wave instabilities depend so much upon the details of the fermiology. Such dependence is the reason why the intertwined electronic ordered states in correlated superconducting compounds are so strongly material dependent. Moreover, precisely due to these distinct intertwined orders \Eq{eff} does not describe a "nearly antiferromagnetic Fermi liquid".  Thus, our approach indicates that many of the anomalous properties of the cuprates and the pnictides may be due to the
fluctuations of the order parameter among the relevant intertwined orders, while the severity of these fluctuations can be material dependent.
We understand that the point of view presented here is much simplified. However, with a goal of identifying concepts that can simply relate strong antiferromagnetic electron-electron interactions, intertwined electronic ordered phases, and strongly correlated superconductivity in distinct material types, this is perhaps a good thing.  We hope that the approach presented here can help to distill  the essence of the unconventional pairing mechanism from the impressive phenomenology of the intertwined phases in present and future strongly correlated high temperature superconductors.

\section{Methods}

We follow the following procedures to determine the favored competing orders from the effective Hamiltonian. Starting with
\be
H_{\rm{eff}}={\sum}'_{\alpha,\v k} \sum_s\epsilon_{\alpha}(\v k) n_{\alpha,s}(\v k)+\sum_{i,j}J_{ij}\vec{S}_i\cdot\vec{S}_j\label{S1}\ee
where $n_{\alpha,s}(\v k)=\psi^\dagger_{\v k,\alpha,s}\psi_{\v k,\alpha,s}$ and $\psi^\dagger_{\v k,\alpha,s}$ creates an electron in the single-particle eigenstate at momentum $\v k$ in band $\alpha$ and with spin $s$. As mentioned in the text ${\sum}'_{\alpha,\v k}$ restricts the sum to single particle eigenstates whose energy is within a thin shell from the fermi energy.

First, we re-express the second term in terms of the band eigen basis:
\be
&&\sum_{i,j}J_{ij}\vec{S}_i\cdot\vec{S}_j
={1\over A}\sum _{\v k,\v p,\v q}\sum_{s_{1,2,3,4}}V_{\v q}(\v k;\v p)\psi^\dagger_{\v k+\v q,s_1}\vec{\s}_{s_1,s_2}\psi_{\v k,s_2}\cdot\psi^\dagger_{\v p-\v q,s_3}\vec{\s}_{s_3,s_4}\psi_{\v p,s_4};\label{ef}\ee where \be &&V_{\v q}(\v k;\v p)=J(\v q) \{\phi_{\alpha(\v k+\v q)}^*(\v k+\v q)\cdot\phi_{\alpha(\v k)}(\v k)\}\{\phi_{\alpha(\v p-\v q)}^*(\v p-\v q)\cdot\phi_{\alpha(\v p)}(\v p)\}.\nonumber
\label{bc}\ee
Here $A$ is the total area, $\phi$ is the band eigen wavefunctions in the orbital basis, and $J(\v q)$ is the Fourier transform of $J_{ij}$. For the copper-based, iron-based, and heavy fermion superconductors $J(\v q)$ is taken to be
an over all coupling strength $J_{\rm eff}$ times the following form factors:
\be &&\cos k_x+\cos k_y ~~({\rm copper-based})\nn &&\cos\theta (\cos k_x+\cos k_y)+\sin\theta (2 \cos k_x\cos k_y) ~~({\rm iron-based})\nn &&\cos k_x+\cos k_y~~({\rm heavy-fermion}),\ee  $J_{\rm eff}$ is a renormalized coupling strength which is a priori unknown. The result for the iron-based superconductors were generated with $\theta=0.3\pi$.
Note in \Eq{ef} we did not keep the band indices. This is because with the restriction to a thin energy shell, momentum actually fixes the band index. In \Eq{bc} the band index, e.g., $\alpha(\v k+\v q)$, is defined to be the the index of the band that is closest to the fermi energy at momentum $\v k+\v q$. If the corresponding single particle state has energy beyond the energy shell, $\phi$ is set to zero. In \Eq{bc} $\phi$ is unity, a 2-component vector, and a  5-component vector for the cuprates, CeCoIn$_5$ and pnictides, respectively. For CeCoIn$_5$ if one decides to include the magnetic interaction between the
f electrons only, one needs to replace $\phi_{\alpha(\v p)}^*(\v p)\cdot\phi_{\alpha(\v q)}(\v q)$ in \Eq{bc} by $\phi_{2,\alpha(\v p)}^*(\v p)\phi_{2,\alpha(\v q)}(\v q)$ where ``2'' labels the f electron Wannier orbitals. The results for CeCoIn$_5$ remain qualitatively unchanged using either formula.

The next step is to decouple \Eq{ef} in the particle-particle (for Cooper pairing) and particle-hole (for charge and spin density wave and Pomeranchuk). The  ``first-instability-mode analysis'' described in section I-III allows us to determine the functional form of the order parameter. However it does not fix the overall magnitude.  Once the functional form  is determined we use the mean-field Hamiltonians described in section I-III to determine the energy gaps, fermi surface distortions, ...etc. The overall magnitude of the order parameter is chosen to yield approximately the same maximal energy gap when each order parameter exists alone. The purpose is to convey the qualitative features not to make quantitative comparative predictions.

\subsection{I. Cooper pairing}

In the particle-particle channel we have focused on the spin singlet Cooper pairing. This leads to the following ``factorization'' of \Eq{ef}:
\be &&H_{\rm{MF}}={\sum}'_{\alpha,\v k} \sum_s\epsilon_{\alpha}(\v k) n_{\alpha,s}(\v k)+{3\over A}\sum _{\v p, \v k}\sum_{a,b} V_{sc}(\v p;\v k)
\Big\{\psi_{-\v k,a}^+\psi_{\v k,b}^+\Delta (\v p)\e_{ab}+\Delta^*(\v k)\e_{ba}\psi_{\v p,a}\psi_{-\v p,b}\nn&&+\Delta^*(\v k)\Delta(\v p)\e_{ab}^2\Big\}.\label{cooper}\ee
Here $a,b$ label the spin and $\e_{\uparrow\downarrow}=-\e_{\downarrow\uparrow}=1$ and $\e_{\uparrow\uparrow}=\e_{\downarrow\downarrow}=0$.
In \Eq{cooper}
\be
V_{sc}(\v p;\v k)&&=J(\v p-\v k)\left(\phi_{\alpha(\v k)}^*(-\v k)\cdot\phi_{\alpha(\v p)}(-\v p)\right)\left(\phi_{\alpha(\v k)}^*(\v k)\cdot\phi_{\alpha(\v p)}(\v p)\right).\ee

We then ``integrate out'' the electrons and keep up to the quadratic terms in $\Delta$'s.  The result is the following free energy form
\be
{1\over A}{\sum}'_{\v k,\v p}\Delta(\v p)K_T(\v p;\v k)\Delta^*(\v k),\ee where
\be
K_T(\v p;\v k)=6 V_{sc}(\v p;\v k)-36\int{d^2q\over(2\pi)^2} V_{sc}(\v p;\v q)\chi_T(\v q)V_{sc}(\v q;\v k),\ee
where the temperature ($T$)-dependent free fermion pair susceptibility is given by
\be
\chi_{T}(\v q)
\propto{1-2f(\e(\v k))\over \e(\v k)}.\ee Here the proportionality constant is un-important for our purposes as it can be absorbed into the unknown $J_{\rm eff}$ (see below).

The leading (sub-leading) gap functions are the eigenfucntions of $M_T(\v q;\v k)=\chi_T(\v q)V_{sc}(\v q;\v k)$ with the largest (second largest) eigenvalue. (The proportionality constant in $\chi_T$ changes all eigenvalues by the same multiplicative constant but not the eigenfunctions.) These are the order parameters which will first (second) become unstable as $J_{\rm eff}$ increases (at a temperature $T$ much less than the thickness of the energy shell). These eigenfunctions are obtained numerically after discretizing the momentum space enclosed by the energy shell (under such discretization $M_T(\v q,\v k)$ becomes a matrix). We diagonalize the $M_T$ matrix then average the eigenfunctions along the direction perpendicular to the fermi surface. This leads to the results presented in the text.

\subsection{II. Charge Density Wave and Pomeranchuk Instability}

Charge density wave and Pomeranchuk instability occur in the spin singlet particle-hole channel. Decoupling \Eq{S1} in this channel leads to the following mean-field Hamiltonian:
 \be
&&H_{\rm{MF}}={\sum}'_{\alpha,\v k} \sum_s\epsilon_{\alpha}(\v k) n_{\alpha,s}(\v k)-{3\over A} \sum _{\v k,\v p,\v Q}V_{cdw}(\v p;\v k)\Big\{\Delta_{\v Q}(\v p)\psi^\dagger_{\v k,a}\psi_{\v k+\v Q,a}+\psi^\dagger_{\v p+\v Q,a}\psi_{\v p,a}\Delta_{\v Q}^*(\v k)\nn&&-2\Delta _{\v Q}(\v p)\Delta_{\v Q}^*(\v k)\Big\}.\label{cdw}\ee
Here 
\be
&&V_{cdw}(\v p;\v k)=J(\v p-\v k)\left(\phi_{\alpha(\v p+\v Q)}^*(\v p+\v Q)\cdot\phi _{\alpha(\v k+\v Q)}(\v k+\v Q)\right)\left(\phi_{\alpha(\v k)}^*(\v k)\cdot\phi_{\alpha(\v p)}(\v p)\right).
\ee
Again, we integrate out the fermions to arrive at the following quadratic free energy form
\be
{1\over A}{\sum}_{\v k,\v p,\v Q}\Delta_{\v Q}(\v p)\tilde{K}_{\v Q,T}(\v p;\v k)\Delta_{\v Q}^*(\v k),\ee where
\be
&&\tilde{K}_{\v Q,T}(\v p;\v k)=6 V_{cdw}(\v k;\v p)-{9}\int {d^2q\over(2\pi)^2}V_{cdw}(\v p;\v q)\tilde{\chi}_{\v Q,T}(\v q)V_{cdw}(\v q;\v k).\ee
Here the free fermion particle-hole susceptibility is given by
\be
\tilde{\chi}_{\v Q,T}(\v q)\propto {f(\e(\v q+\v Q))-f(\e(\v q))\over\e(\v q)-\e(\v q+\v Q)}.\label{phchi}\ee The leading order parameter is the eigenfunction of $\tilde{M}_{\v Q,T}(\v q,\v k)=\tilde{\chi}_{\v Q,T}(\v q)V_{cdw}(\v q;\v k)$ with the largest  eigenvalue. Here we have to search both the ordering wavevector $\v Q$ as well as the leading form factor. This is again achieved numerically after discretizing the momentum space within the energy shell and diagonalize the resulting matrix $\tilde{M}_{\v Q,T}$. As in section I we perform an average of the eigenvector along the direction perpendicular to the fermi surface, which leads to the results presented in the text.

The Pomerahnchuk distortion is determined as the leading order parameter in the $\v Q\ra 0$ limit of $M_{\v Q,T}$. In our calculation we always find both a $\v Q=0$ and $\v Q\ne 0$ instabilities.

\subsection{III. Spin Density Wave}

 Spin density wave is a spin triplet particle-hole instability. Decoupling \Eq{S1} in this channel leads to the following mean-field Hamiltonian:
 \be
H_{\rm{MF}}&&={\sum}'_{\alpha,\v k} \sum_s\epsilon_{\alpha}(\v k) n_{\alpha,s}(\v k)
+{1\over A}\sum _{\v p,\v k,\v Q} V_{sdw}(\v p;\v k)\Big\{{\v m}_{\v Q}(\v p)\cdot\psi^\dagger_{\v k,c}\vec{\sigma }_{cd}\psi_{\v k+\v Q,d}
+{\v m}^*_{\v Q}(\v k)\cdot\psi^\dagger_{\v p+\v Q,a}\vec{\sigma }_{ab}\psi_{\v p,b}\nn&&-\v m_{\v Q} (\v p)\cdot {\v m}^*_{\v Q} (\v k)
\Big\},
\label{sdw}\ee
where
\be &&V_{sdw}(\v p;\v k)=J(\v Q)\left(\phi_{\alpha(\v p+\v Q)}^*(\v p+\v Q)\cdot\phi_{\alpha(\v p)}(\v p)\right)\left(\phi_{\alpha(\v k)}^*(\v k)\cdot\phi _{\alpha(\v k+\v Q)}(\v k+\v Q)\right)\nn&&+{1\over 2}J(\v p-\v k)\left(\phi_{\alpha(\v p+\v Q)}^*(\v p+\v Q)\cdot\phi_{\alpha(\v k+\v Q)}(\v k+\v Q)\right)\left(\phi _{\alpha(\v k)}^*(\v k)\cdot\phi_{\alpha(\v p)}(\v p)\right).\ee

Like in section I and II, we integrate out the fermions. The resulting quadratic free energy form read
\be
{1\over A}{\sum}_{\v k,\v p}K_{sdw,\v Q,T}(\v p;\v k)\v m(\v p)\cdot\v m(\v k),\ee where
\be
&&K_{sdw,\v Q,T}(\v p;\v k)=-V_{sdw}(\v k;\v p)-2\int {d^2q\over(2\pi)^2}V_{sdw}(\v p;\v q)\tilde{\chi}_{\v Q,T}(\v q)V_{sdw}(\v q;\v k).\ee
Here the free fermion particle-hole susceptibility is given by \Eq{phchi}. The leading order parameter is the eigenfunction of $\tilde{M}_{\v Q,T}(\v q,\v k)=\tilde{\chi}_{\v Q,T}(\v q)V_{sdw}(\v q;\v k)$ with the minimum eigenvalue. As in section I and II we search the leading order parameter numerically after discretizing the momentum space within the energy shell.

\subsection{IV. The Pseudogap of the Cuprates}

Fig.4 of the main text is generated by superposing the order parameter terms in \Eq{cooper} (SC), \Eq{cdw}(CDW) and \Eq{sdw}(SDW) to form a grand mean-field Hamiltonian.
 If we include all necessary components the super-vector with antiferromagnetic, superconducting, and charge density wave order as components will have 3+2+2+2=9 components. (The last 2+2 is the number of components of the charge density wave, associated with, e.g., the $(\pm\delta,0)$ and $(0,\pm\delta)$ order.) This is too complex to handle and impossible to present the results. We simplify the situation to a super-vector with only three components. The first component is the superconducting order. Here we restrict the phase of the superconducting order parameter to be real. The second component is the charge density wave order. Here we choose a bi-directional charge density wave with the two fundamental density wave order in-phase coexist, and we pin the over all (sliding) phase of the order parameter. The third component is the antiferromagnetism. Here we restrict the order parameter to point in a particular, say the z, direction. We also rescale the order parameters so that when exists alone, each component gives rise to an approximately equal maximal gap. This leads to the following mean-field Hamiltonian
 \be &&H_{\rm{MF}}={\sum}'_{\alpha,\v k} \sum_s\epsilon_{\alpha}(\v k) n_{\alpha,s}(\v k)+n_1\Big\{{1\over A}\sum _{\v p,\v k}  V_{sdw}(\v p;\v k)\Big[f_{sdw,\v Q_s}(\v p)\psi^\dagger_{\v k,c}\sigma^z_{cd}\psi_{\v k+\v Q_s,d}
\nn&&+f^*_{sdw,\v Q_s}(\v k)\psi^\dagger_{\v p+\v Q_s,a}\sigma^z_{ab}\psi_{\v p,b}\Big]
 +n_2\Big\{{1\over A}\sum _{\v p, \v k}\sum_{a,b}3\e_{ab}V_{sc}(\v p;\v k)\Big[f_{sc}(\v p)\psi_{-\v k,a}^+\psi_{\v k,b}^+ +f^*_{sc}(\v k)\psi_{\v p,b}\psi_{-\v p,a}\Big]\Big\}\nn&&+n_3\Big\{{1\over A}\sum _{\v k,\v p,\v Q_c}(-3)V_{cdw}(\v p;\v k)\Big[f_{cdw,\v Q_c}(\v p)\psi^\dagger_{\v k,a}\psi_{\v k+\v Q_c,a}+f_{cdw,\v Q_c}^*(\v k)\psi^\dagger_{\v p+\v Q_c,a}\psi_{\v p,a}\Big]\Big\}.\ee  Here $\v Q_s=(\pi,\pi)$, $\v Q_c=(\pm\delta,0),(0,\pm\delta)$, and $f_{sdw,\v Q_s},f_{sc},f_{cdw\v Q_c}$ are the form factors of the leading order parameters determined in section I-III, properly scaled to produce a similar maximum gap when each order parameter exists alone. The $n_1,n_2,n_3$ are the components of the super-vector shown in Fig.4.
 In general for incommensurate $\delta$ the above mean-field Hamiltonian couples infinite many $\v k$ points together. The result presented in the main text is obtained by truncating this infinite set to the following 10 elements set $\{\v k, \v k\pm (\delta,0), \v k\pm (0,\delta), \v k+(\pi,\pi), \v k+(\pi,\pi)\pm (\delta,0), \v k+(\pi,\pi)\pm (0,\delta)\}$. This leads to a $20\times 20$ Nambu matrix for each $\v k$. This matrix is diagonalized numerically to determine the energy gap. What's plotted in Fig.4 is the minimum energy gap among all $\v k$ (within the energy thin shell) for each direction normal to the fermi surface.

\begin{acknowledgments}
We thank S.A Kivelson for a most useful discussion on the $\v Q=0$ instabilities in the cuprates. We are also grateful to D.K. Morr, M. Norman, and  S. Sachdev for helpful discussions and communications. JCD is supported by the Center for Emergent Superconductivity, an Energy Frontier Research Center, headquartered at Brookhaven National Laboratory and funded by the U.S. Department of Energy, under DE-2009-BNL-PM015; by the Office of Naval Research under Award N00014-13-1-0047. DHL is supported by DOE Office
of Basic Energy Sciences, Division of Materials Science, under Material Theory program,
DE-AC02-05CH11231.
\end{acknowledgments}

\newpage

\noindent{{\bf{Figure Cations}}}\\

\noindent{{\bf Figure 1.}} {\bf Schematic phase diagram of unconventional superconductors}
Starting from a robust phase of commensurate antiferromagnetism (AF) a control parameter, such as carrier density or pressure, is varied so that the critical temperature T$_{\rm AF}$ of the AF phase diminishes. Eventually, an unconventional superconducting (SC) phase appears at higher values of the control parameter and its critical temperature T$_c$ is usually 'dome' shaped. The intervening gray region is where the antiferromagnetic phase and the superconducting phase connect. It is here that the intertwined phases (IP) of electronic matter have typically been discovered. The characteristics of the intertwined phases are highly distinctive to each system, as is the precise way (e.g. first order, coexistence, quantum critical ... etc.) that the AF-SC connection occurs. By contrast, the appearance of unconventional  SC phase upon suppression of an AF state is virtually universal.
\\

\noindent{{\bf Figure 2.}} {\bf Fermi Surface and Unconventional Superconducting States of Cuprates}
(a) 	The cuprate first Brillouin Zone (BZ), within which all the momentum-space ($\v k$-space) electronic states of the system are described when not in the AF state. It spans a range $-\pi/a<k_x\le\pi/a$; $-\pi/a<k_y\le\pi/a$ where $a$ is the unit cell dimension. The dimensions of the BZ in Fig.2(a) are in units of $\pi /a$. The model Fermi surface of the cuprates, constructed using a tight-binding single band model with 1st ($t$), 2nd ($t'$) and 3rd ($t''$) neighbor hopping, where $t'/t = 0.3$ and $t''/t = 0.2$, is shown.
(b) 	This Fermi surface exhibits eight special "hot spots" (the red dots) where the AF BZ (dashed lines) crosses it. They appear to play a leading role in the interplay of intertwined phases and superconductivity\cite{KFujita,Sachdev2}. The black and gray arrows are the wavevectors of the leading and subleading charge density wave instability.
 (c) 	The leading spin-singlet superconducting gap function derived from \Eq{eff} (Methods Section I). The hatch size is proportional to the magnitude and the color indicates the sign (red:-, blue:+).
(d) 	The sub-leading singlet superconducting gap function derived from \Eq{eff} (Methods Section I). The hatch size is proportional to the magnitude and the color indicates the sign (red:-, blue:+). The gap functions in panel (c) and (d) are well described by $\cos k_x - \cos k_y$ and $\cos k_x +\cos k_y$, respectively.
\\

\noindent{{\bf Figure 3.}}{\bf The $\v Q=0$ intertwined particle-hole instabilities of the cuprates} (Methods section II) (a) The order parameter of the leading $\v Q = 0$ (Pomeranchuk) instability and the associated Fermi surface distortion of cuprates derived from Eq. 1 (Methods Section II).  Here the hatch size is proportional to the magnitude and the color indicates the sign (red:-, blue:+), and the dashed line marks the un-distorted Fermi surface. This instability breaks the 90$^\circ$ rotation symmetry and leads to nematicity. (b,c) The order parameter of a degenerate pair of subleading $\v Q=0$ instabilities and the associated Fermi surface distortions. Because the distorted Fermi surfaces are not $\v k\leftrightarrow -\v k$ symmetric, these instabilities lead to time reversal symmetry breaking.
\\

\noindent{{\bf Figure 4.}} {\bf The ground state current of the T-breaking $\v Q=0$ particle-hole instabilities in the cuprates} (a) The ground state current distribution associated with the order parameters in Fig.3(b) and Fig.3(c).  (b) The ground state current distribution produced by the symmetric and antisymmetric linear combination of the order parameters in Fig.3(b) and Fig.3(c). In (a,b) the thickness of the arrow is proportional to the magnitude of the current.
\\

\noindent{{\bf Figure 5.}} {\bf The leading intertwined $\v Q\ne 0$ particle-hole instabilities of the cuprates}(Methods Section II). The ordering wavevector is $(\pm\delta, 0)$ in panel (a,b) and $(0,\pm\delta)$ in panel (c,d). The black and gray arrows in Fig.2b are approximately the ordering wavevectors of the leading and sub-leading CDW instabilities, respectively. (e) The energy gap of CDW produced by the equal amplitude superposition of the charge density wave depicted in panel (a)-(d). The phase of the superposition is taken to be $+,+,-,-$, hence corresponds to a $d$-wave symmetry. (The energy gap associated with the $+,+,+,+$ superposition of panel  (a)-(d) is similar.) The charge density wave gap plotted in panel  (e) is defined as the minimum energy gap of the mean-field Hamiltonian in (Method section II) along the momentum cut normal to the Fermi surface but within the energy thin shell.
The smallness of this combined order parameter near the nodes can give rise to the effect of "Fermi arcs".
\\

\noindent{{\bf Figure 6.}}{\bf Intertwined Instabilities and the Cuprate Pseudogap}
We represent the three instabilities contained in \Eq{eff} (AF, SC and CDW) using a super-vector that represent the combination of the three order parameters.  The admixture of AF, SC and CDW phases can then be indicated by using the location of the super-vector on a sphere. In each panel from (a) to (e) the direction of the super-vector is shown as the red arrow on the order parameter sphere on the right of the same panel. The gap shown on the left of each panel is the minimum energy gap of the mean-field Hamiltonian in (Method section IV) along the momentum cut normal to the Fermi surface but within the energy thin shell.
The size of the hatch is proportional to the value of the single particle energy gap.
\\

\noindent{{\bf Figure 7.}} {\bf Model Fermi Surface and Superconducting States of Iron-Pnictides}
(a) The pnictide first Brillouin Zone (BZ) when not in the AF state. It spans a range $-\pi/a<k_x\le k_x/a; -\pi/a<k_y\le\pi/a$ where $a$ is the dimension of a unit cell containing only one Fe atom (we neglect the effects on unit cell definition of the out of plane As atoms). Our model Fermi surface of iron-pnictides using a five-band tight-binding model is shown as five closed contours, two red (outline the electron pocket) and three blue (outline the hole pocket),
(b) The leading spin singlet superconducting gap function derived from \Eq{eff}. (Methods Section I). The symmetry is $S\pm$. The hatch size is proportional to the magnitude and the color indicates the sign (red:-, blue:+).
(c) The subleading singlet superconducting gap function derived from \Eq{eff} (Methods Section I). The symmetry is $d_{x^2-y^2}$. The hatch size is proportional to the magnitude and the color indicates the sign (red:-, blue:+). The result in panel (b) and (c) are obtained using $J_2/J_1 = \tan 0.3\pi$.
\\

\noindent{{\bf Figure 8.}} {\bf Leading Intertwined Instabilities of Iron-Pnictides}
(a) 	The leading $\v Q=0$ instability and the associated Fermi surface distortions in the iron-based superconductors derived from \Eq{eff} (Methods Section II). The hatch size is proportional to the magnitude and the color indicates sign (blue:+, red:-). The dashed lines mark the un-distorted Fermi surface. The area of both electron and hole pockets shrink so that the total charge density is kept constant. This "distortion" does not break any symmetry, hence is difficult to pin down.
(b) 	The sub-leading $\v Q=0$ instability and the associated Fermi surface distortion derived from \Eq{eff} (Methods Section II). This distortion breaks the $90^\circ$ rotation symmetry and couples to the "stripe-like" unidirectional antiferromagnetic correlation strongly. Although the instability in panel (b) is subleading at the quadratic level of the Landau free energy expansion, it can become leading once the cubic coupling with the antiferromagnetic order parameter is taken into account.
(c) 	The effect of the Fermi surface distortion in panel (b) on the orbital occupation $n_{xz}(\v k)-n_{yz}(\v k)$.
\\

\noindent{{\bf Figure 9.}} {\bf Fermi Surface and Unconventional Superconducting States of the Heavy-Fermion {\bf compound CeCoIn$_5$}.} (a) The first Brillouin zone and Fermi surface associated with a two-band band structure in Ref.\cite{Allan}. The BZ spans a range $-\pi/a < k_x \le \pi/a;-\pi/a < k_y\le\pi/a$ where $a$ is the dimension of a unit
cell. The leading (b) and sub-leading (c) spin singlet superconducting gap functions. The leading gap function has d$_{x^2-y^2}$ symmetry and the sub-leading one has extended S symmetry. In panels (b) and (c) the hatch size is proportional to the magnitude of the gap and the color indicates the sign (red:-, blue:+). (d) The fermi surface and hot spots (the pink dots) of CeCoIn$_5$.
\\

\noindent{{\bf Figure 10.}} {\bf The intertwined $\v Q=0$ particle-hole Instabilities of CeCoIn$_5$.} (a) The leading Pomeranchuk instability and the associated Fermi surface distortion. The hatch size is proportional to the magnitude of the order parameter and the color indicates the sign (blue:+ red:-). The dashed line marks the un-distorted Fermi surface. This Fermi surface distortion leads to the breaking of the 90$^\circ$ rotation symmetry. (b,c) The degenerate pair of subleading Pomeranchuk instabilities and their Fermi surface distortions. In both panels the distorted Fermi surfaces do not respect the $\v k\leftrightarrow -\v k$ symmetry. Consequently time reversal symmetry is broken.
\\

\noindent{{\bf Figure 11.}} {\bf The leading intertwined $\v Q\ne 0$ particle-hole Instabilities of CeCoIn$_5$.}  (a-d) The leading charge density wave order parameter. (a,b) The ordering wavevectors are $\pm(0.56\pi,0.26\pi)$. (c,d) The ordering wavevectors are $\pm(0.26\pi,0.56\pi)$. (e) The energy gap produced by the in-phase coexistence of order parameters in panels (a)-(d). In this figure the hatch size is proportional to the magnitude and the color indicates the sign (red:-, black and blue:+).






\begin{figure}
\includegraphics[scale=.7]{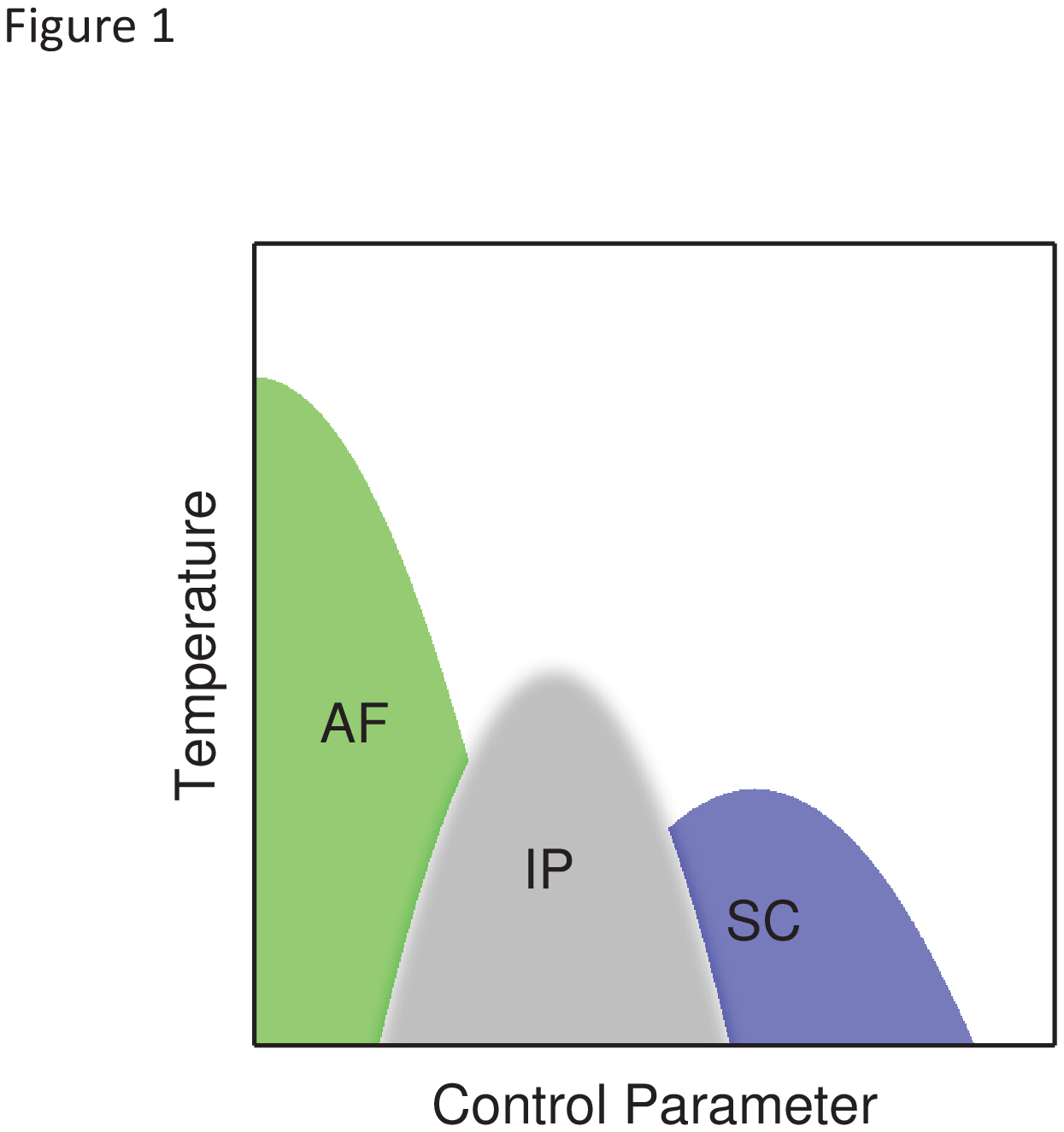}
\caption{}
\end{figure}

\clearpage

\begin{figure}
\includegraphics[scale=.7]{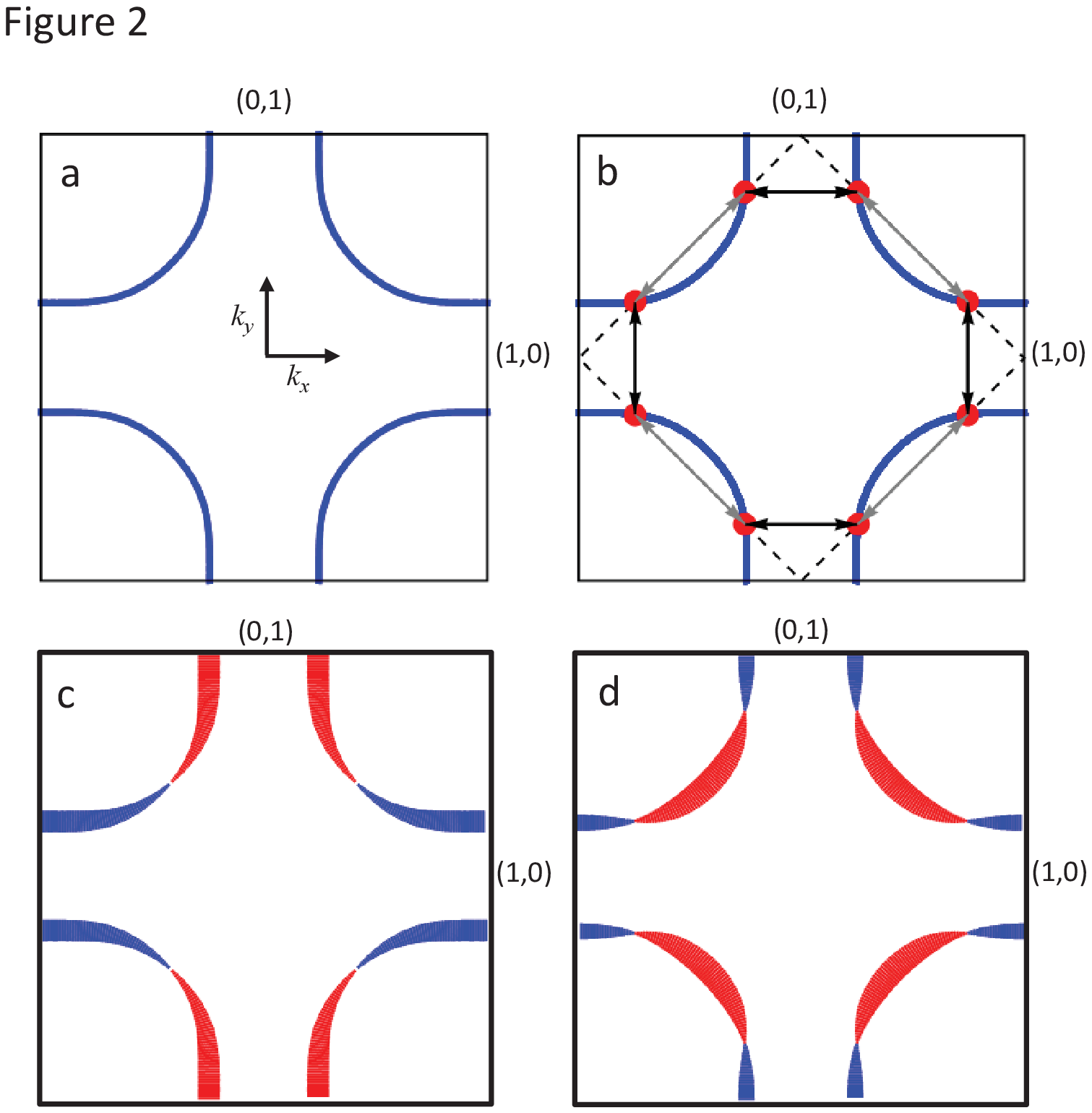}
\caption{}
\end{figure}

\clearpage

\begin{figure}
\includegraphics[scale=.7]{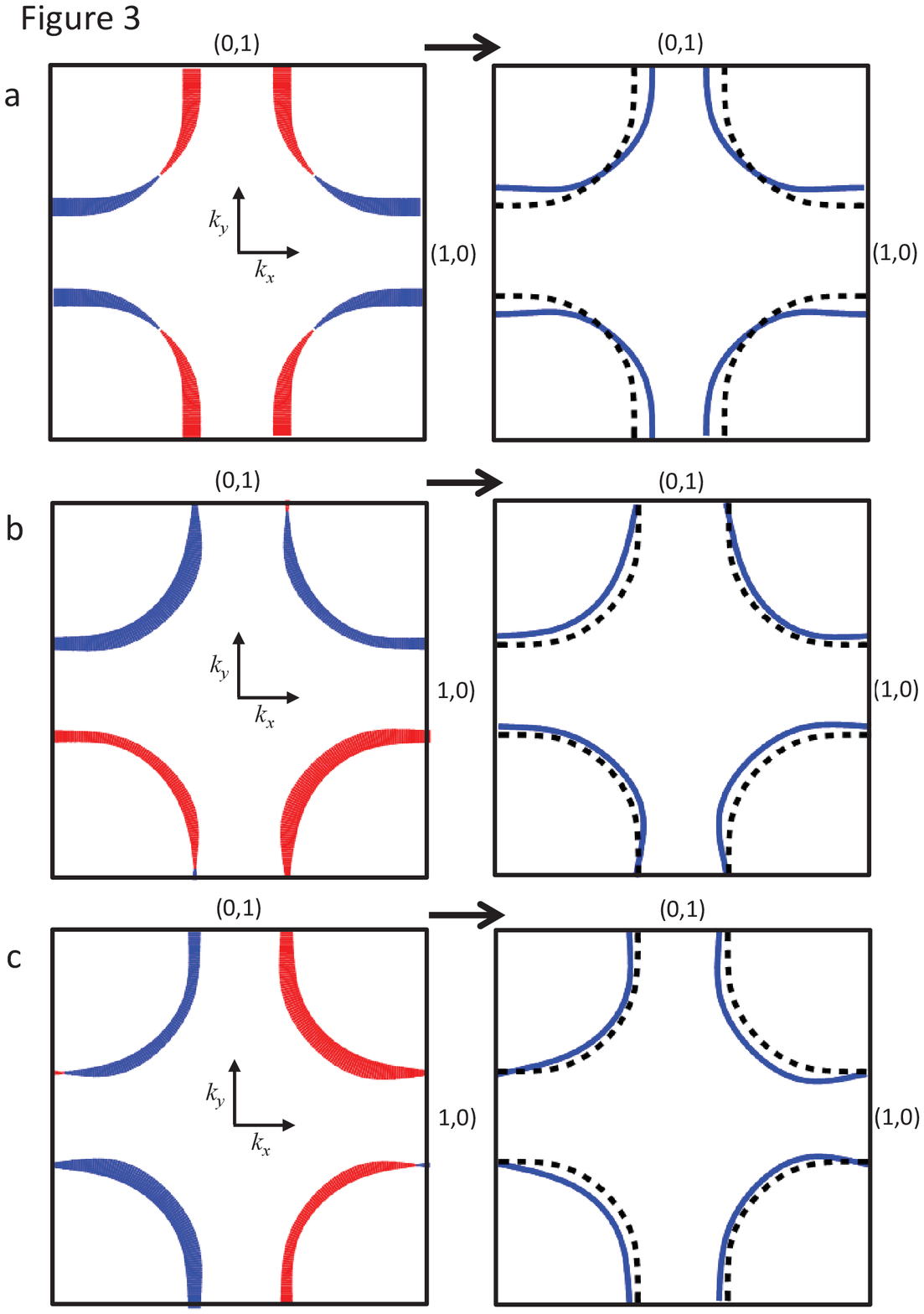}
\caption{}
\end{figure}

\clearpage

\begin{figure}
\includegraphics[scale=.7]{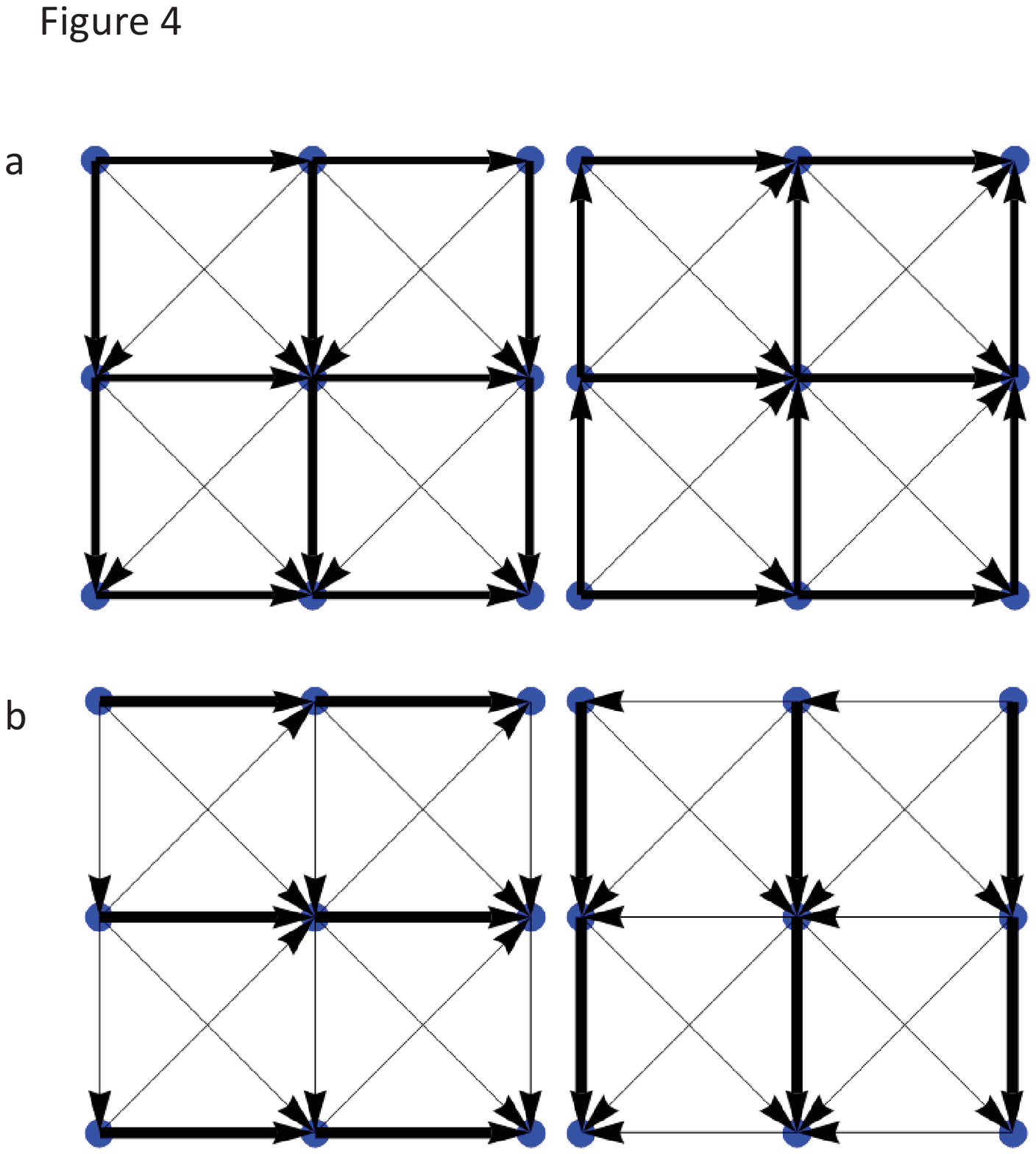}
\caption{}
\end{figure}

\clearpage

\begin{figure}
\includegraphics[scale=.7]{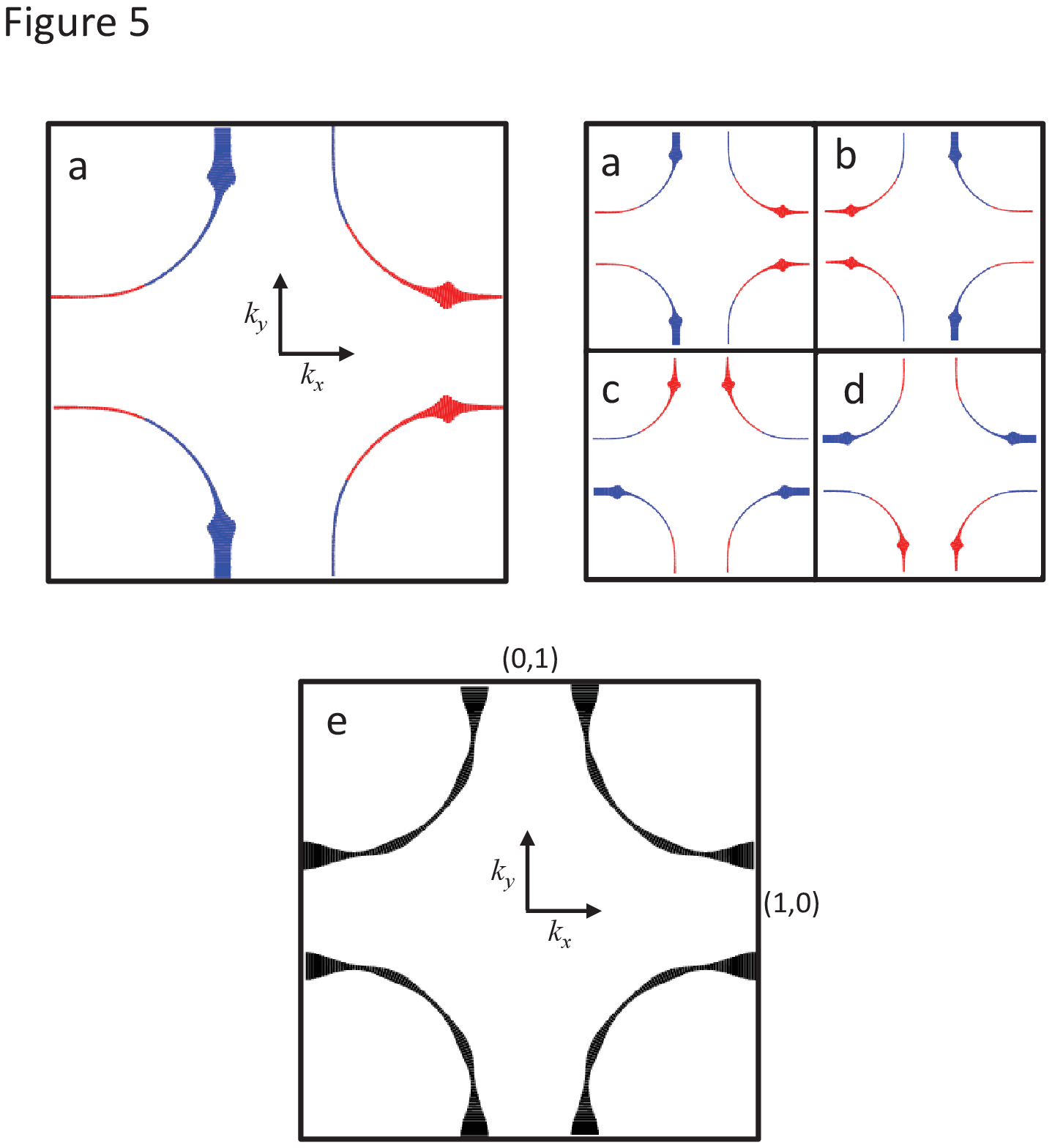}
\caption{}
\end{figure}

\clearpage

\begin{figure}
\includegraphics[scale=.7]{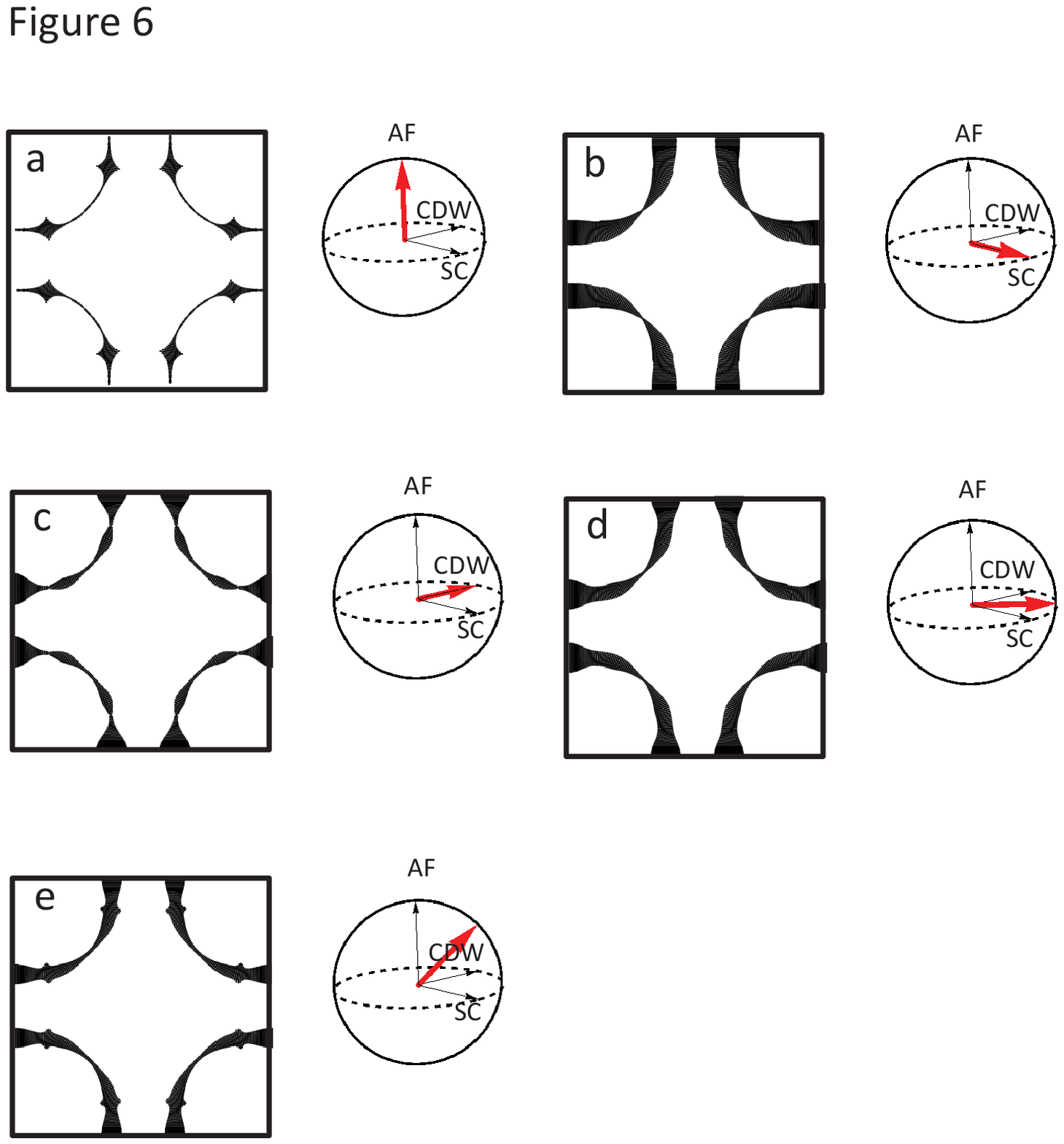}
\caption{}
\end{figure}

\clearpage

\begin{figure}
\includegraphics[scale=.7]{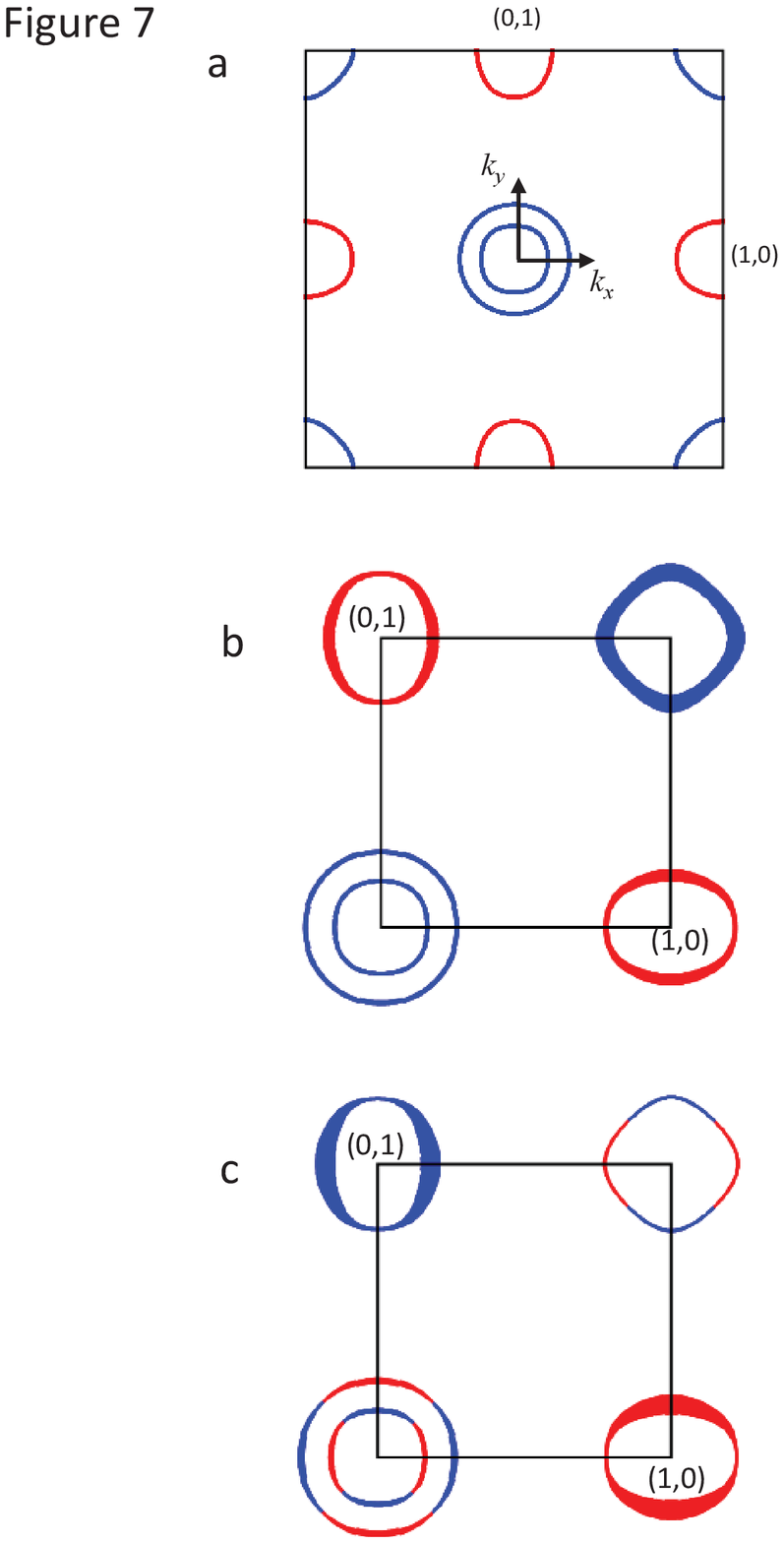}
\caption{}
\end{figure}

\clearpage

\begin{figure}
\includegraphics[scale=.7]{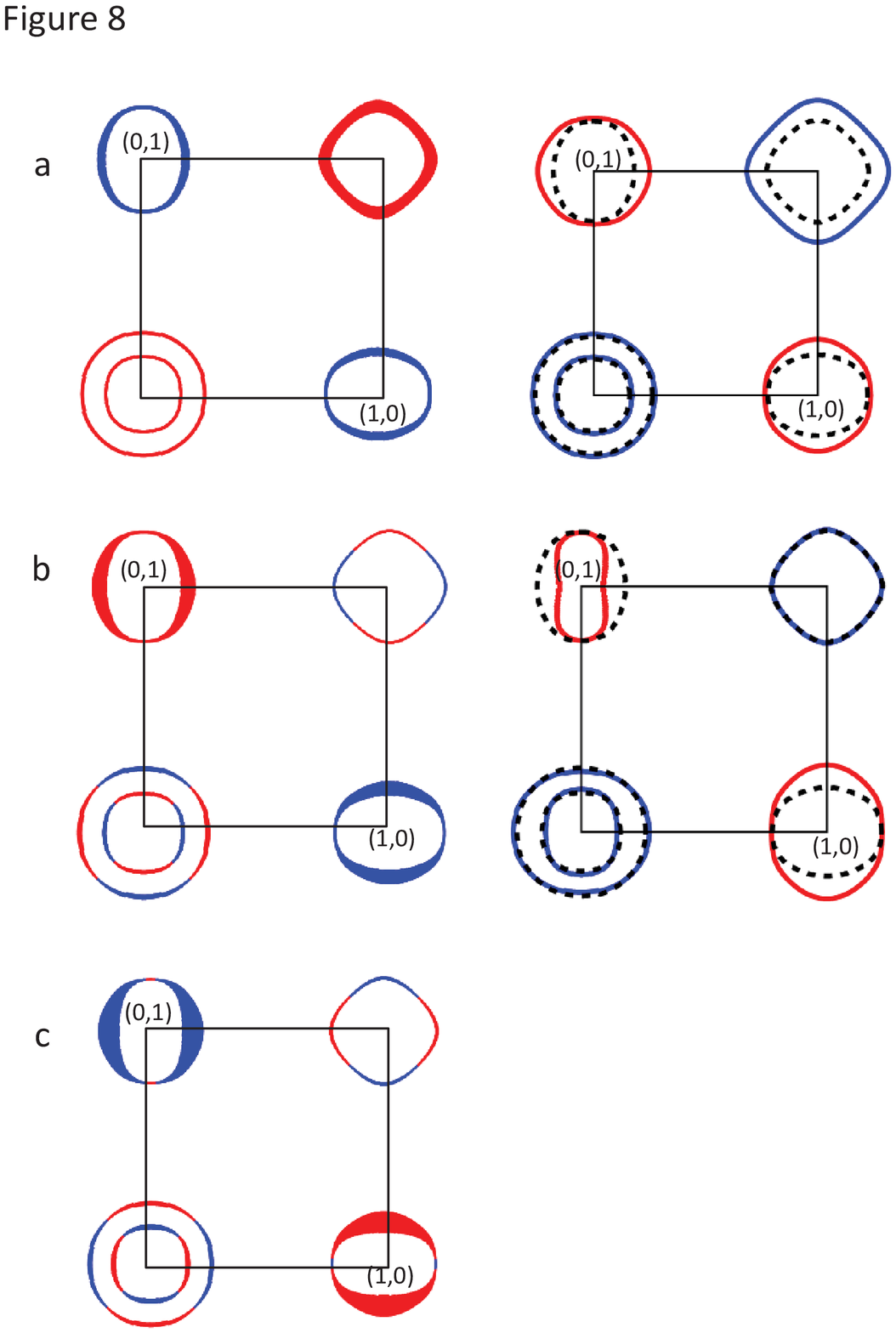}
\caption{}
\end{figure}

\clearpage

\begin{figure}
\includegraphics[scale=.7]{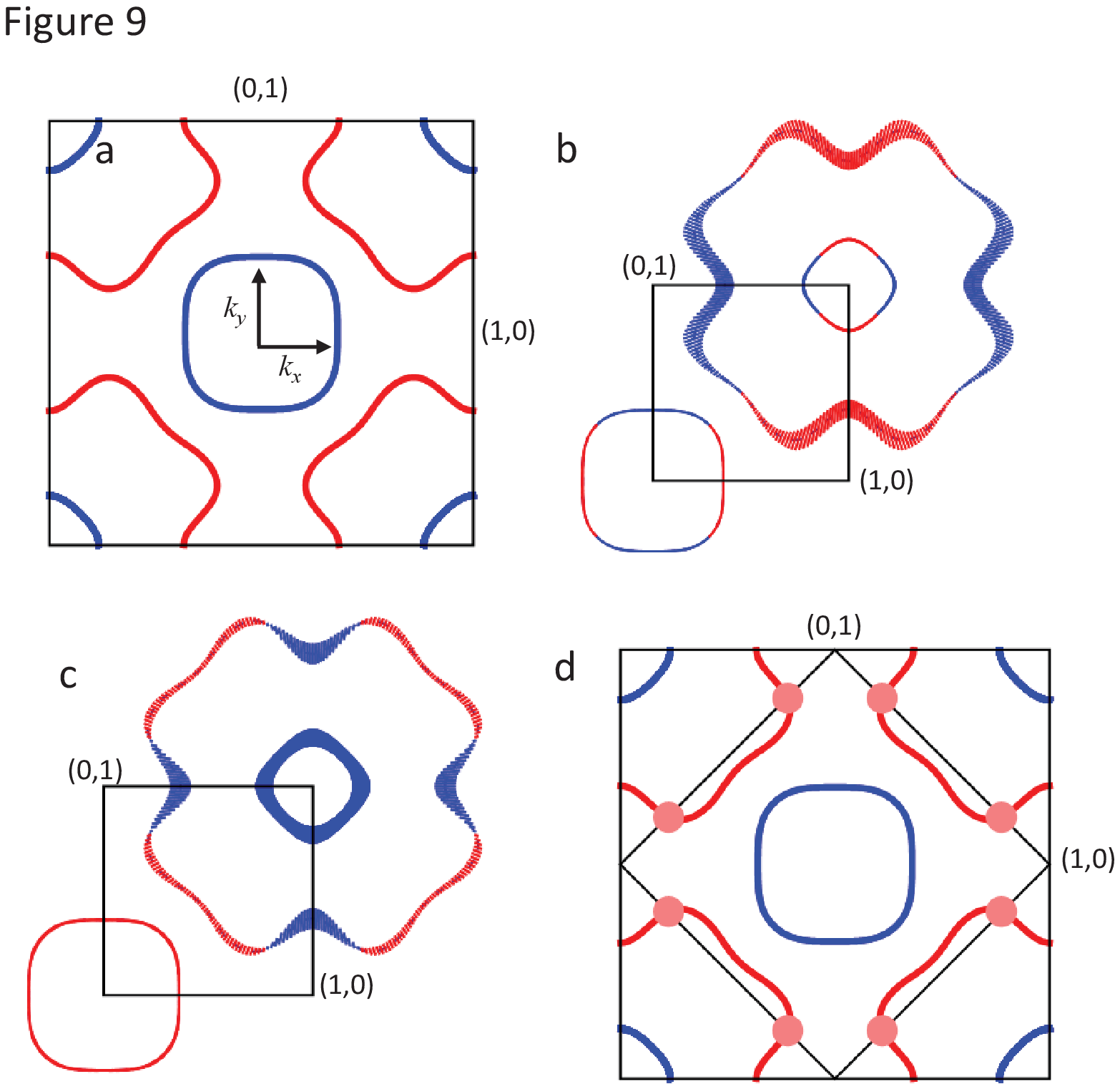}
\caption{}
\end{figure}

\clearpage

\begin{figure}
\includegraphics[scale=.7]{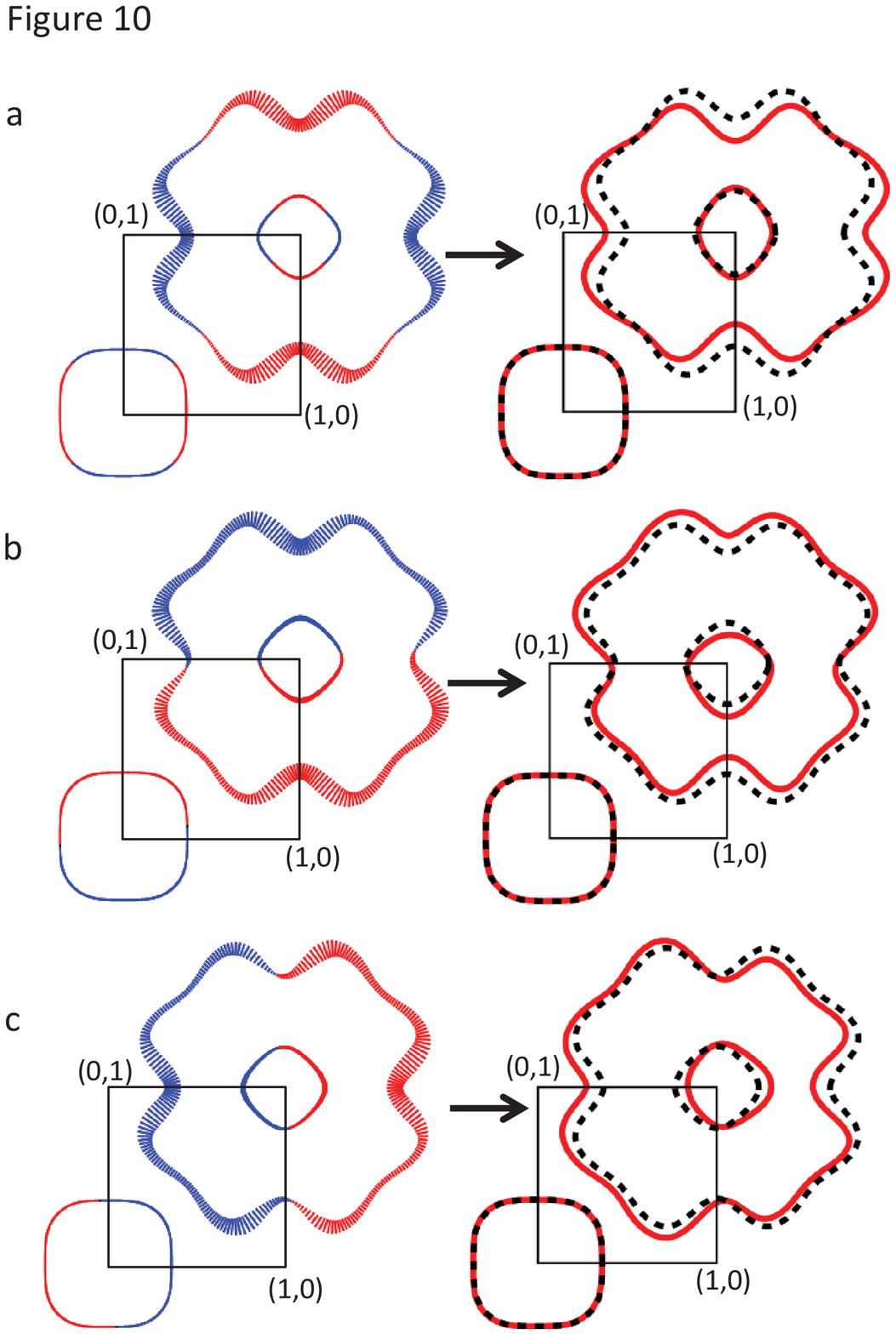}
\caption{}
\end{figure}

\clearpage

\begin{figure}
\includegraphics[scale=.7]{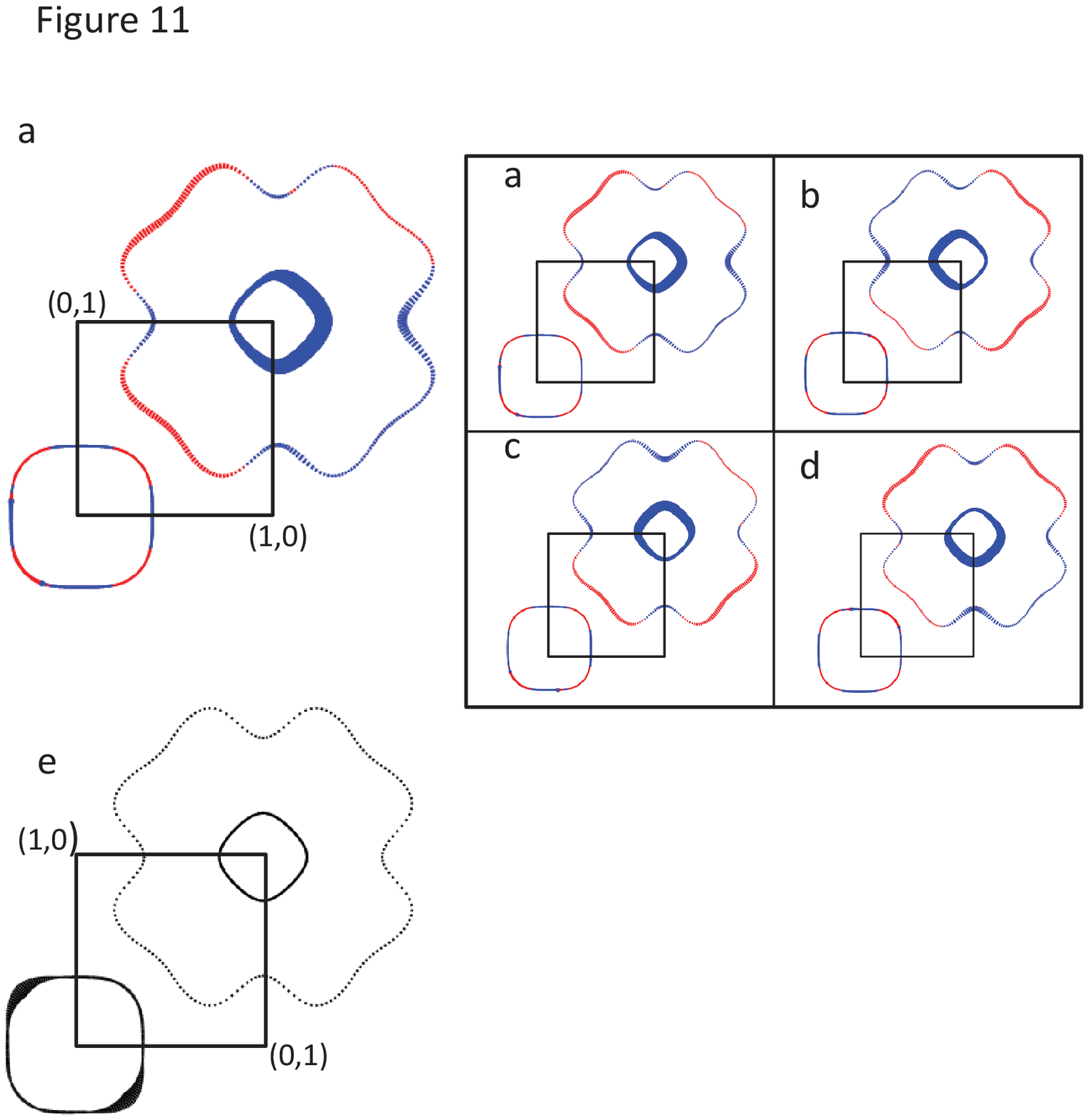}
\caption{}
\end{figure}

\end{document}